\documentclass[aip,jap,12pt]{revtex4-2}

\usepackage{xcolor}
\usepackage{graphicx}
\usepackage{float}
\usepackage{xr}

\begin{document}

\title{Oxidation-resilient structural modifications in Nickel-functionalized 3D-graphene for hydrogen storage applications} 

\author{Filippo Fincato}
\affiliation{Istituto Nanoscienze–CNR, NEST-Scuola Normale Superiore, Piazza San Silvestro 12, 56127 Pisa, Italy}

\author{Ylea Vlamidis}
\affiliation{Istituto Nanoscienze–CNR, NEST-Scuola Normale Superiore, Piazza San Silvestro 12, 56127 Pisa, Italy}
\affiliation{Department of Physical Science, Earth, and Environment, University of Siena, Via Roma 56, 53100 Siena, Italy}

\author{Markus Leitgeb}
\affiliation{Institute of Sensor and Actuator Systems, TU Wien, 1040 Vienna, Austria}

\author{Georg Pfusterschmied}
\affiliation{Institute of Sensor and Actuator Systems, TU Wien, 1040 Vienna, Austria}

\author{Ulrich Schmid}
\affiliation{Institute of Sensor and Actuator Systems, TU Wien, 1040 Vienna, Austria}

\author{Valentina Zannier}
\affiliation{Istituto Nanoscienze–CNR, NEST-Scuola Normale Superiore, Piazza San Silvestro 12, 56127 Pisa, Italy}

\author{Silvia Rubini}
\affiliation{CNR - Istituto Officina dei Materiali (IOM), S.S. 14 Km 163.5, 34149 Trieste, Italy}

\author{Stefan Heun}
\affiliation{Istituto Nanoscienze–CNR, NEST-Scuola Normale Superiore, Piazza San Silvestro 12, 56127 Pisa, Italy}

\author{Stefano Veronesi}
\affiliation{Istituto Nanoscienze–CNR, NEST-Scuola Normale Superiore, Piazza San Silvestro 12, 56127 Pisa, Italy}

\date{\today}

\begin{abstract}
\section*{Abstract}
Porous materials represent a versatile solution for several applications. Indeed, the recent development of a new material, named 3D-Graphene, which combines the exceptional characteristics of graphene with a three-dimensional structure, opens perspectives for applications where a high surface-to-volume ratio is beneficial. In this study, we explore the functionalization of 3D-Graphene with nickel (Ni)-nanoparticles as a strategy to enhance hydrogen storage capabilities, and we assess the influence of the NPs on hydrogen uptake and oxidation resilience. The morphology and structural properties of pristine and Ni-functionalized samples were characterized using Scanning Electron Microscopy. Additionally, X-ray Photoelectron Spectroscopy was employed to analyze the surface chemical composition of the functionalized samples. Samples have been hydrogenated supplying molecular or atomic hydrogen, and hydrogen storage performance was assessed through Thermal Desorption Spectroscopy. Afterwards, oxidation effects were systematically studied by exposing the samples to atmospheric oxygen, followed by further hydrogenation experiments. Our results indicate that Ni functionalization influences both hydrogen adsorption and oxidation behavior, with potential implications for improving the stability of the material, especially for hydrogen storage applications.

\end{abstract}

\maketitle

\section{Introduction}

The increasing global demand for clean and efficient energy storage solutions has driven extensive research into advanced materials for hydrogen storage \cite{Rampai2024, Yang2023}. Indeed, hydrogen, a promising energy carrier \cite{Guilbert2021, Staffell2019}, requires efficient storage methods to enable its large-scale adoption in hydrogen-fueled energy systems \cite{Gunathilake2024, Khalili2025}. While conventional hydrogen storage approaches, such as compressed gas and cryogenic liquid storage, pose challenges in terms of safety, volumetric density, and energy efficiency \cite{Gardiner2009}, solid-state hydrogen storage has emerged as a viable alternative \cite{Yang2023, Martelli2010}. Graphene-based materials have attracted significant attention for this purpose due to their high surface area, tunable adsorption properties, and their potential for functionalization \cite{Farrukh2021, Chen2022}.

In particular, three-dimensional graphene structures (3D-Graphene or 3DG) have been developed to overcome the limitations of monolayer graphene \cite{Veronesi2022}. 3DG provides a porous network that enhances molecular adsorption while maintaining the intrinsic properties of graphene \cite{Veronesi2022, Macili2023}. The three-dimensional structure serves several applications such as catalysis, high performance electrodes, drug delivery, water treatment, and gas detection and storage \cite{Kadja2022,Ahmad2023,Mabrouk2019,Memetova2022,Veronesi2025}. 
Recent studies have demonstrated that the functionalization of graphene with metals can further improve its hydrogen storage capacity, primarily through catalytic effects and spillover mechanisms \cite{Chen2022, Ferbel2024}. Moreover, porous materials can be efficiently functionalized with nanoparticles \cite{Pompei2024}. Nickel is of particular interest, due to its hydrogen dissociation properties and its ability to enhance adsorption kinetics when incorporated into graphene-based structures \cite{Kim2005, Zhou2016}. However, the stability of Ni-functionalized 3D-Graphene under oxidative environments remains a critical aspect that requires further investigation.

Exposure to oxygen can significantly alter the surface chemistry of graphene and metal nanoparticles, impacting their adsorption properties \cite{Modi2019, Liu2020, Anas2025}. Understanding the role of oxidation in the functionalization process and its impact on hydrogen uptake is essential for assessing the feasibility of these materials for applications. Indeed, commercial-grade hydrogen contains up to 5 ppm of oxygen (EN 17124:2022), inducing an oxidation with time of any material chosen as storage medium. In this work, we investigate the intercalation of Ni nanoparticles within 3D-Graphene and systematically evaluate the effect on both hydrogen storage performance and oxidation resilience.
We employ Scanning Electron Microscopy (SEM) to analyze the morphology of pristine and Ni-functionalized 3D-Graphene samples. Additionally, X-ray Photoelectron Spectroscopy (XPS) is used to probe oxidation-induced modifications, while Thermal Desorption Spectroscopy (TDS) provides insight into hydrogen uptake efficiency before and after exposure to oxygen. By comparing pristine and functionalized samples under controlled conditions, we aim to clarify the interplay between Ni intercalation, hydrogen adsorption, and oxidative effects in 3D-Graphene. The findings of this study contribute to a deeper understanding of metal-functionalized graphene for hydrogen storage applications and provide new insight into strategies for enhancing material stability in oxidative environments.

\section{Materials and Methods}

\subsection{Sample Preparation}

Porous silicon carbide (pSiC) substrates were obtained via metal-assisted photochemical etching and photoelectrochemical etching (MAPCE–PECE) \cite{Leitgeb2013}. The 4H-SiC wafers used in this work were subjected to a two-step photoelectrochemical etching process. The first step, Metal-Assisted Photochemical Etching (MAPCE), begins with the deposition of 300 nm Pt pads onto the wafer, followed by UV illumination in an electrochemical cell containing an HF-based oxidizing solution. Pt acts as the cathode, while the exposed SiC surface serves as the anode. The process induces localized oxidation and dissolution, forming a porous surface layer.
This porous structure enables the subsequent Photo-Electrochemical Etching (PECE), performed in a separate HF-based solution under UV illumination. The sample is positioned between two compartments of the PECE cell, acting as a barrier between the electrodes. When a bias is applied, anodic oxidation and HF-assisted dissolution occur on the face adjacent to the cathode, resulting in finer porosification. 

Graphene was then formed by thermal annealing of the porous SiC in ultra-high vacuum at high temperature, around 1400 \textdegree C, following the procedure described in \cite{Veronesi2022}. The graphene quality was verified by Raman spectroscopy. 

Functionalization was achieved by introducing nickel nanoparticles (Ni NPs) --- synthesized according to the method described in \cite{Vlamidis2024} --- via immersion of the 3DG samples in a Ni NPs colloidal suspension for 48 h, following the procedure outlined in \cite{Pompei2024}. In brief, a solution of cetyltrimethylammonium bromide (CTAB) (12 mM) and Ni(OAc)$_2$ (25 mM) in milliQ water was cooled down to $\sim 1$ \textdegree C   in an ice bath for 30 minutes. Then it was vigorously stirred, adding 200 $\mu$L of sodium borohydride (20 mg/mL in milliQ water) in 5 aliquots, followed by 2 minutes of additional stirring. After the metal reduction, the solution is centrifuged at 13000 rpm for 10 minutes to remove the larger NPs precipitate. The smaller nanoparticles (diameter $<15$ nm), suspended in the supernatant, are collected by dilution of the solution with ethanol and washed several times with water by centrifugation and re-suspension steps, to remove excess reagents.

\subsection{Scanning Electron Microscopy (SEM)}

For SEM measurements conducted on functionalized 3DG samples, a FEG-SEM Merlin from ZEISS was used. Samples were fixed on the sample holder using carbon tape to acquire top-view and cross sectional images, employing both secondary electrons (SE) and back-scattered electrons (BSE) in mixed imaging to create mass contrast imaging, which is useful for identifying Ni-rich regions.

\subsection{X-ray Photoelectron Spectroscopy (XPS)}

XPS measurements were performed using a Surface Science Instrument SSX-100-301 spectrometer operating with an Al K$\alpha$ X-ray source with an overall energy resolution of 0.9 eV.  Specctra were aquired at room temperature on Ni-functionalized 3DG samples as a function of annealing temperature, through \textit{in situ} thermal annealings, from room temperature up to 800 \textdegree C. The analysis focused on the C 1s, O 1s, Ni 2p, and Si 2p core levels to monitor chemical state evolution and potential changes in the Ni–3DG interface. Fitting of the spectra is performed after a Shirley background subtraction by deconvolution of the peaks in s- or p- components with a Gaussian-Lorentzian lineshape.

\subsection{Thermal Desorption Spectroscopy (TDS)}

All hydrogenation and desorption experiments were performed in the same UHV chamber, maintaining a base pressure of approximately \(1 \times 10^{-10}\) mbar. Samples are prepared by degassing at 700 \textdegree C overnight. Before the hydrogenation experiments, samples were annealed at 800 \textdegree C or 950 \textdegree C for 10 minutes. The hydrogenation experiments are performed exposing the samples to molecular or atomic deuterium. Deuterium was used to increase the signal-to-noise ratio, and in this work hydrogen and deuterium are used as synonyms. Samples were exposed to hydrogen for 5 minutes at \(1 \times 10^{-7}\) mbar prior to TDS. Desorption spectra were acquired under a constant heating rate of 4\,K\,s\(^{-1}\), ramping up to 750 \textdegree C. Temperature measurements were collected simultaneously via both a thermocouple and an optical pyrometer, ensuring accurate control and monitoring of the sample temperature during desorption. Desorption peaks were fitted using OriginPro software after baseline subtraction to zero. The experimental curves were first smoothed with a 15-point adjacent averaging (AAv) filter, and individual peaks were fitted with Gaussian functions using the Levenberg–Marquardt algorithm. Peak centers were initially identified through an unconstrained preliminary fit and subsequently used as constraints in the final fitting to ensure consistency and reproducibility across spectra.

\section{Results and Discussion}

\begin{figure}[t]
    \includegraphics[width=\columnwidth]{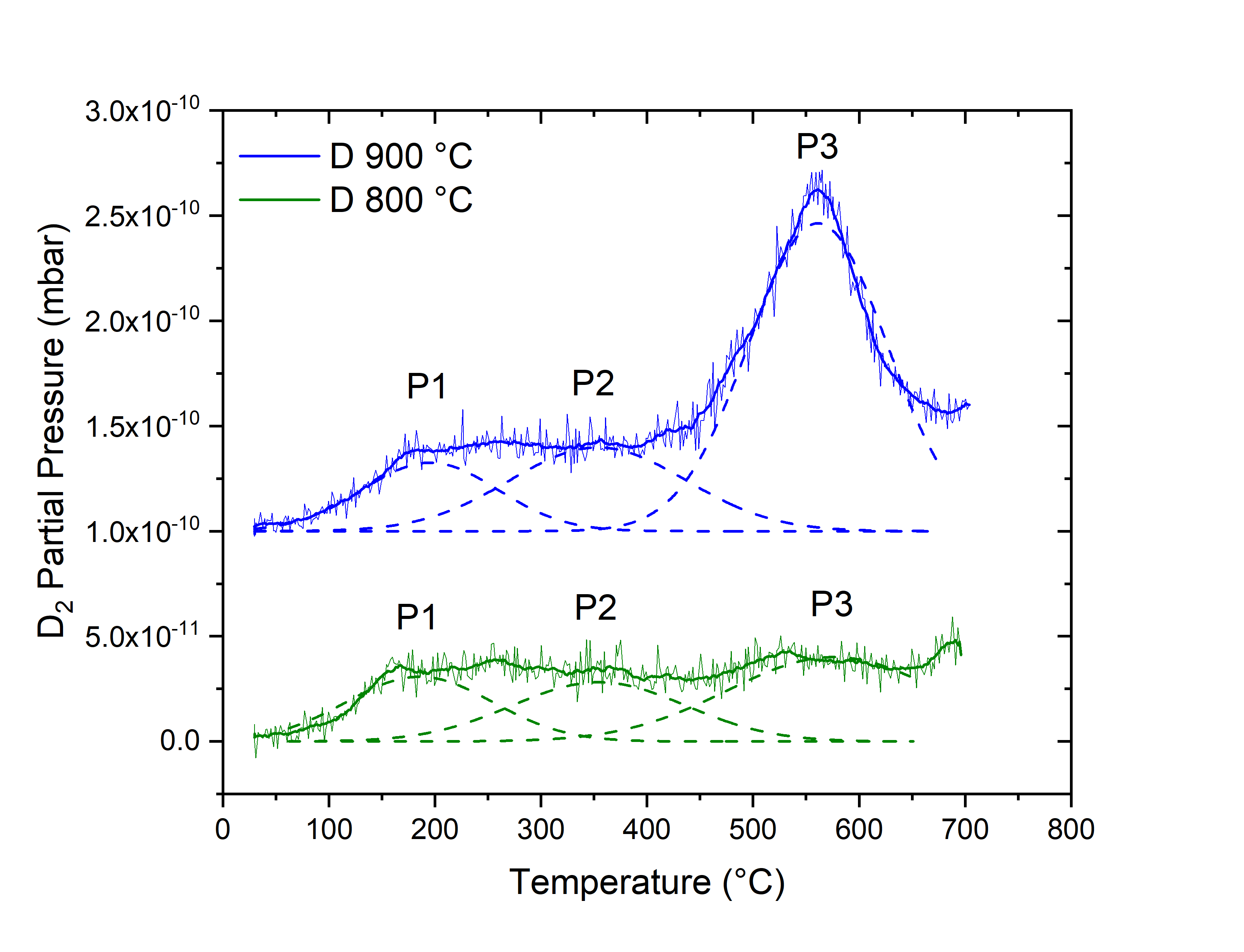}
    \caption{TDS spectra of pristine 3D-Graphene for two annealing conditions: 800 °C annealing (green) and 900°C annealing (blue), both for 10 minutes. Samples were exposed to hydrogen for 5 minutes at \(1 \times 10^{-7}\) mbar prior to TDS, performed with constant heating rate of 4\,K\,s\(^{-1}\). The blue curve was shifted by 1.0$\times 10^{-10}$ mbar for better visualization.}
    \label{fig:1}
\end{figure}

The hydrogen storage evaluation of the samples starts from the pristine 3D-Graphene, which represents our control sample. Figure~\ref{fig:1} shows the TDS spectra of a pristine 3D-Graphene sample after exposure to atomic deuterium. The green curve refers to a sample which was annealed at 800 \textdegree C for 10 minutes before hydrogenation, and which displays a moderate desorption signal spread over a broad temperature range. Previuos experimental work on 3DG samples showed three desorption peaks in the range $150-550$ \textdegree C \cite{Macili2023}. The spectra were thus deconvoluted by a Gaussian fit into three distinct desorption peaks, P1 centered at 185 \textdegree C, P2 centered at 350 \textdegree C, and P3 centered at 550 \textdegree C. 

After a second annealing at 900 \textdegree C for 10 minutes followed by a hydrogenation (blue curve in Figure~\ref{fig:1}), the sample exhibits a markedly higher desorption signal with a pronounced peak centered around 550 \textdegree C. These observations suggest that higher annealing temperatures may enhance the availability of active sites able to adsorb hydrogen, with a mechanism unfolding between 800 \textdegree C and 900 \textdegree C. This behavior suggests that the Ni-functionalized samples should be subjected to similar annealings, to maximise the hydrogen uptake. 

\begin{figure}[t]
    \includegraphics[width=\textwidth]{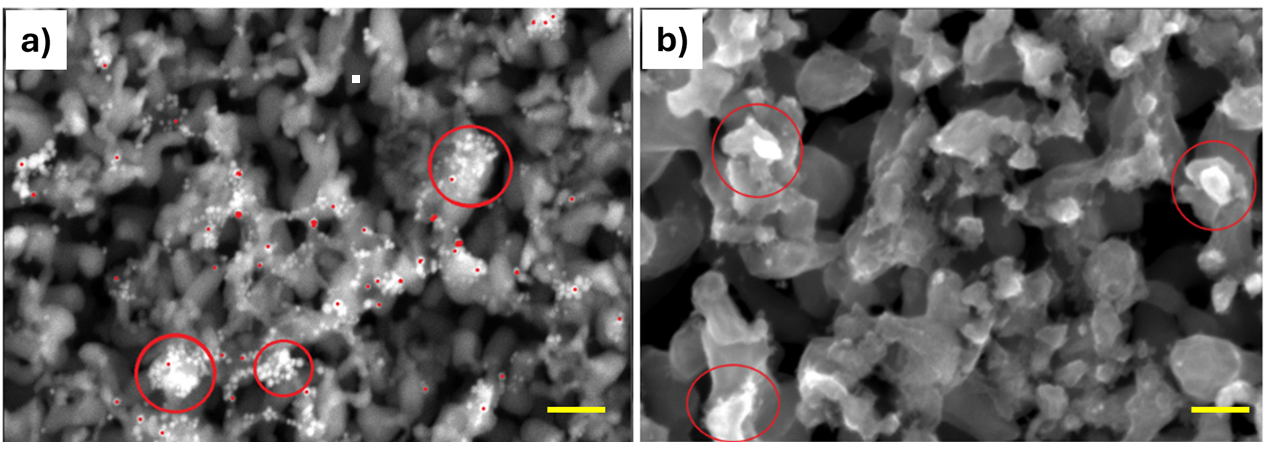}
    \caption{Morphological changes observed in Ni-functionalized 3D-Graphene upon progressive heat treatments. (a) SEM Image of a Ni-3DG sample after functionalization but before thermal treatment: NiNPs are distinguishable as singles (red dots) and clusters (red circles). (b) SEM image of the same Ni-3DG sample after 900 \textdegree C annealing: although individual Ni NPs are no longer visible, Ni-rich regions (red circles) are still detectable through mass-contrast BSE imaging, highlighting compositional variations within the matrix. Scale bars: a) 200 nm; b) 100 nm.}
    \label{fig:3}
\end{figure}

Focusing on the Ni NPs, Figure~\ref{fig:3} illustrates the morphological changes observed in Ni-functionalized 3D-Graphene upon progressive heat treatments. In the SEM image reported in Figure~\ref{fig:3}a (pre-annealing), nickel nanoparticles are clearly visible as discrete entities (either individual particles or small clusters). After annealing at 900 \textdegree C, the SEM image reported in Figure~\ref{fig:3}b shows that nanoparticles are no longer distinguishable and cover extended surface areas, likely due to local melting and enhanced mobility of nickel at the nanoscale, despite its substantially higher bulk melting point (1455 \textdegree C) \cite{LANL}.

\begin{figure}[H]    
    \includegraphics[width=0.9\textwidth]{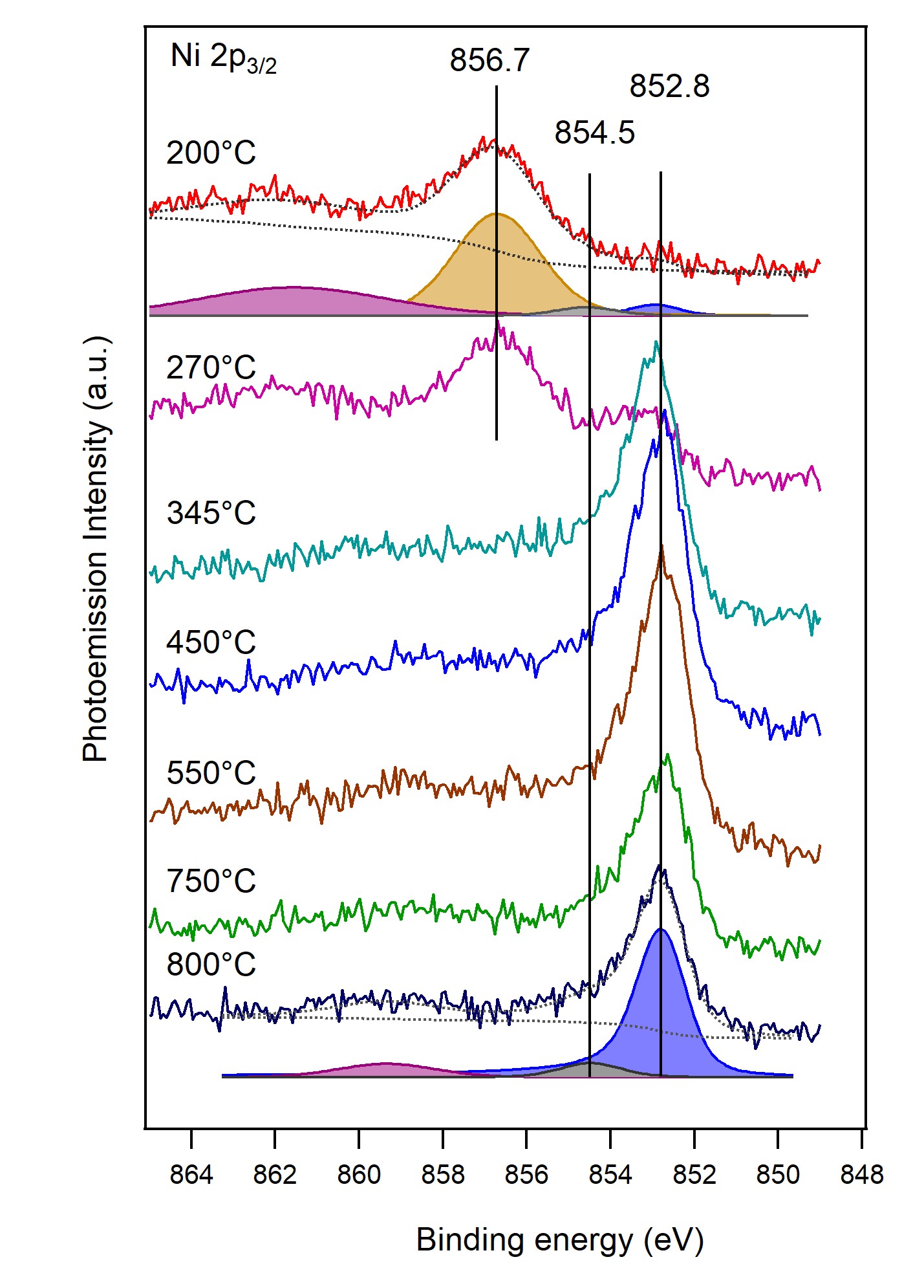}
    \caption{Photoemission spectra of Ni 2p$_{3/2}$ for a Ni-functionalized 3D-Graphene sample recorded after annealing at increasing temperatures, as indicated in the Figure. The spectra have been shifted vertically to allow an easier comparison. The detailed line-shape analysis is shown for the lowest and the highest annealing temperatures and is discussed in the text.}
    \label{fig:4}
\end{figure}

Information about the surface chemistry of the Ni-functionalized 3D-Graphene samples is given by the XPS analysis. Figure~\ref{fig:4} shows the Ni 2p$_{3/2}$ photoemission spectra acquired after successive annealing steps from 200 \textdegree C to 800 \textdegree C. A detailed line-shape analysis is reported for the lowest and the highest annealing temperatures. At 200 \textdegree C, the spectrum is dominated by a broad component centered at 856.7 eV, assigned to Ni(OH)$_2$-like species at the surface of the NPs, likely originating from the CTAB capping procedure and exposure to ambient conditions. As the annealing temperature is increased and the CTAB capping layer desorbs, the metallic Ni 2p$_{3/2}$ component at 852.8 eV becomes progressively dominant. A weaker NiO$_{\textrm{x}}$ contribution at 854.5 eV \cite{Grzelczak2008}, together with the high-binding-energy satellite structure, is detectable. The overall line shape remains essentially unchanged from 450 \textdegree C to 800 \textdegree C. This behavior suggests that no significant chemical transformation of the Ni nanoparticles, such as silicide formation, occurs upon annealing in that temperature range. However, the intensity of the main photoemission feature (i.e., the metallic component) increases from 345 \textdegree C to 450 \textdegree C and subsequently decreases at higher annealing temperatures (see the SI).

\begin{figure}[H]
    \includegraphics[width=0.8\textwidth]{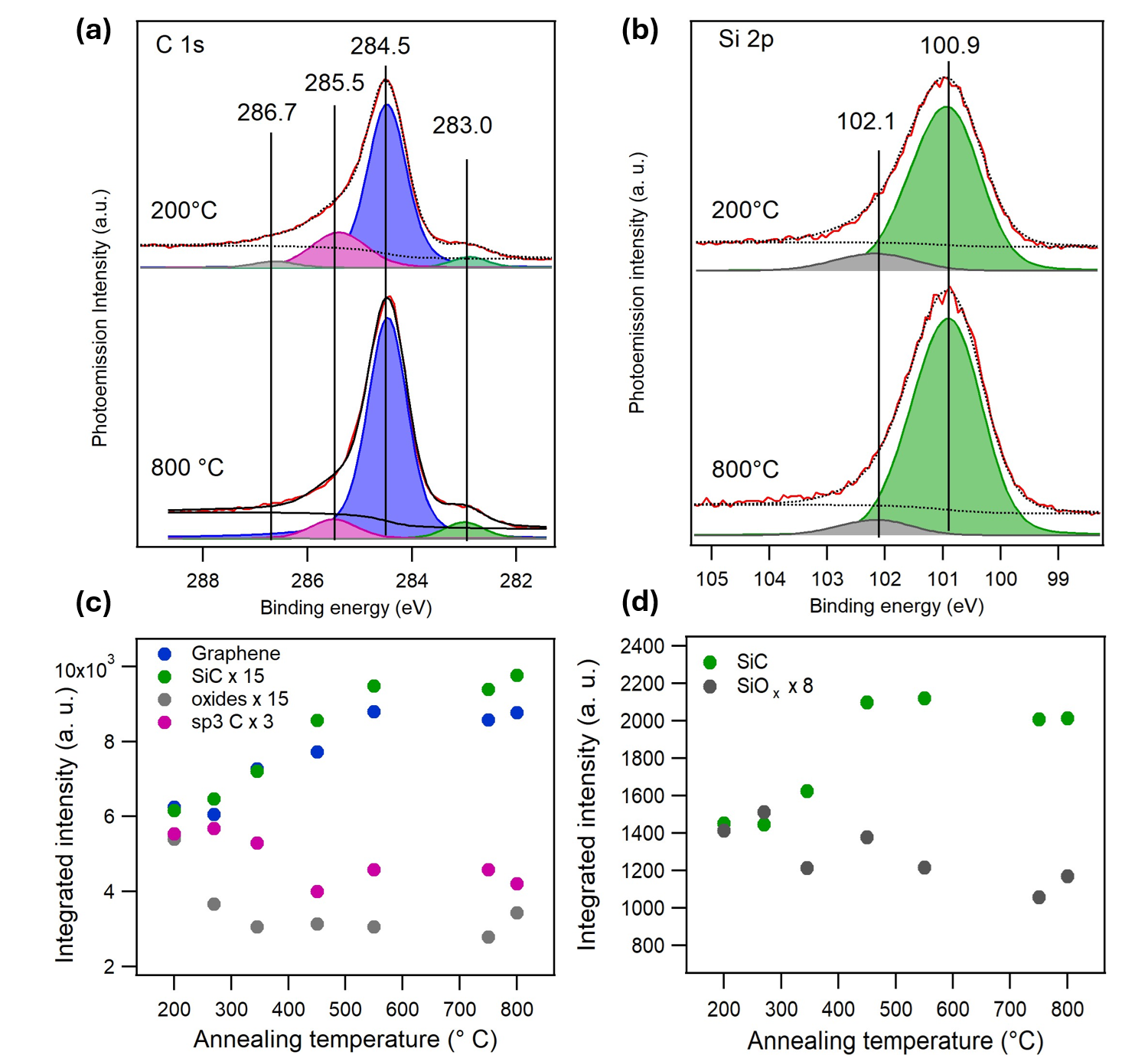}
    \caption{(a) Photoemission spectra of C 1s for a Ni-functionalized 3D-Graphene sample recorded after annealing at 200 \textdegree C and at 800 \textdegree C. The line shape analysis reveals the presence of the Graphene and SiC components at 284.5 eV and 283.0 eV, respectively, together with a weak contribution from oxides (286.7 eV) and sp$^3$ C (285.5 eV). (b) Photoemission spectra of Si 2p for the same sample recorded after annealing at 200 \textdegree C and at 800 \textdegree C. The line shape has been fitted with spin-orbit doublets ($\Delta_{\textrm{so}}$ = 0.63 eV). The analysis displays the presence of the major SiC component with center of mass at 100.9 eV and a minor component at 102.1 eV originating likely from SiO$_{\textrm{x}}$. (c) Integrated intensities of the C 1s components, obtained by line shape analysis of the photoemission spectra, as a function of annealing temperature. (d) Integrated intensity of the Si 2p components, obtained by line shape analysis of the photoemission spectra, as a function of annealing temperature.}
    \label{fig:4a}
\end{figure}

In Figure~\ref{fig:4a}(a) and (b) the C 1s and Si 2p photoemission spectra, respectively, are reported, both recorded after annealing at 200  \textdegree C and at 800 \textdegree C, to highlight the subtle changes induced by the thermal treatment and to elucidate the evolution of the system upon annealing. The C 1s emission is dominated by the sp$^2$ graphene component at 284.5 eV. At higher binding energy, the contributions of interface-related and/or defective (sp$^3$) carbon (at 285.5 eV) and of oxidized species (C-O, C-O-C, at 286.7 eV) are observed, while at lower binding energy (at 283.0 eV) the SiC component is detected \cite{Pompei2024}. 

The Si 2p emission has been fitted with spin-orbit doublets ($\Delta_{\textrm{so}}$ = 0.63 eV). At all temperatures, the spectra display a major SiC component with center of mass at 100.9 eV, and a minor higher-binding-energy component at 102.1 eV originating likely from SiO$_{\textrm{x}}$ \cite{Pompei2024}.

For both core levels, the evolution of the integrated intensity of the above-mentioned components is reported in the panels (c) and (d) of Figure~\ref{fig:4a}. Upon annealing at increasing temperatures, the Graphene contribution to the C 1s core level and the SiC contribution to the photoemission intensities of both core levels increase. The SiO$_{\textrm{x}}$ component decreases slightly, and also the C oxides and sp$^3$ components decrease with annealing temperature. The increase in the emission intensity of the graphene and SiC components in the C 1s spectra, together with the increase of the SiC component in Si 2p and the decrease of the metallic Ni 2p$_{3/2}$ signal, is consistent with an increase in the exposed Gr/SiC area following the clustering of the Ni nanoparticles observed by SEM and/or with Ni intercalation. Indeed, the intercalation of Ni under the graphene layer is documented in the literature \cite{Huang2013}, also after deposition of Ni NPs onto epitaxial graphene \cite{Vlamidis2025}, suggesting that this phenomenon occurs also here.

\begin{figure}[H]
    \includegraphics[width=0.8\columnwidth]{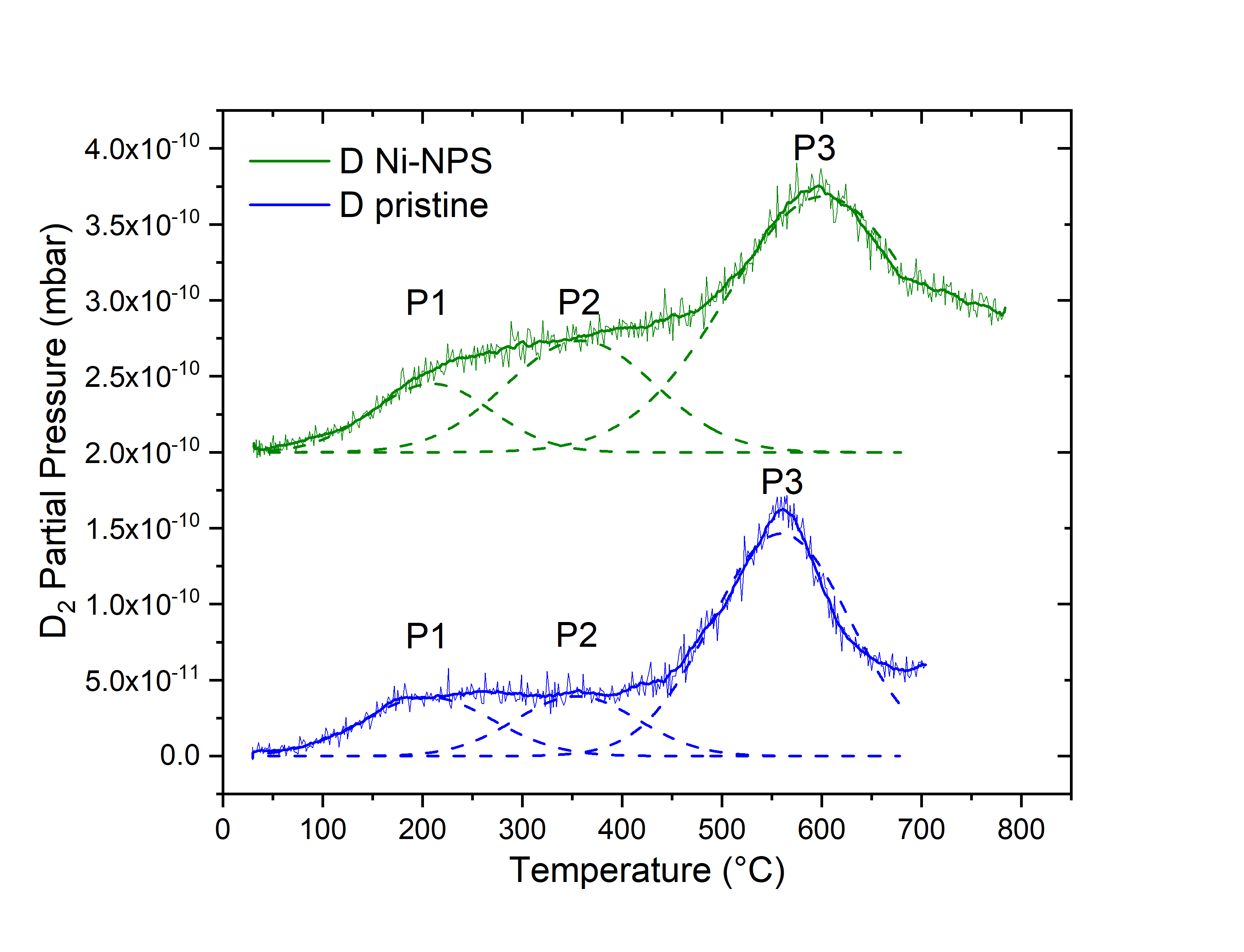}
    \caption{Comparison of TDS spectra for pristine 3D-Graphene (blue) and Ni-functionalized 3D-Graphene (green) both annealed at 900 \textdegree C prior to hydrogenation. Single peak profiles (P1, P2, and P3) were deconvoluted by a Gaussian fit and are reported by dashed lines. Samples were exposed to hydrogen for 5 minutes at \(1 \times 10^{-7}\) mbar prior to TDS, performed with constant heating rate of 4\,K\,s\(^{-1}\). The green curve was shifted by 2.0$\times 10^{-10}$ mbar for better visualization.}
    \label{fig:2}
\end{figure}

After the \textit{ex-situ} functionalization with Ni NPs \cite{Vlamidis2024}, the sample utilized for the hydrogenation experiments shown in Figure~\ref{fig:1} was again loaded in the experimental chamber and prepared for hydrogenation. In the first heating step, the sample was annealed for several hours at 350 \textdegree C, which is the degradation temperature of the capping layer of the Ni NPs \cite{Vlamidis2024}. Successively, the sample was annealed at 900 \textdegree C for 10 min. Figure~\ref{fig:2} reports the TDS spectra of pristine 3D-Graphene and Ni-functionalized 3D-Graphene, both annealed at 900 \textdegree C prior to  exposure to atomic deuterium. 

\begin{figure}[H]
    \includegraphics[width=\columnwidth]{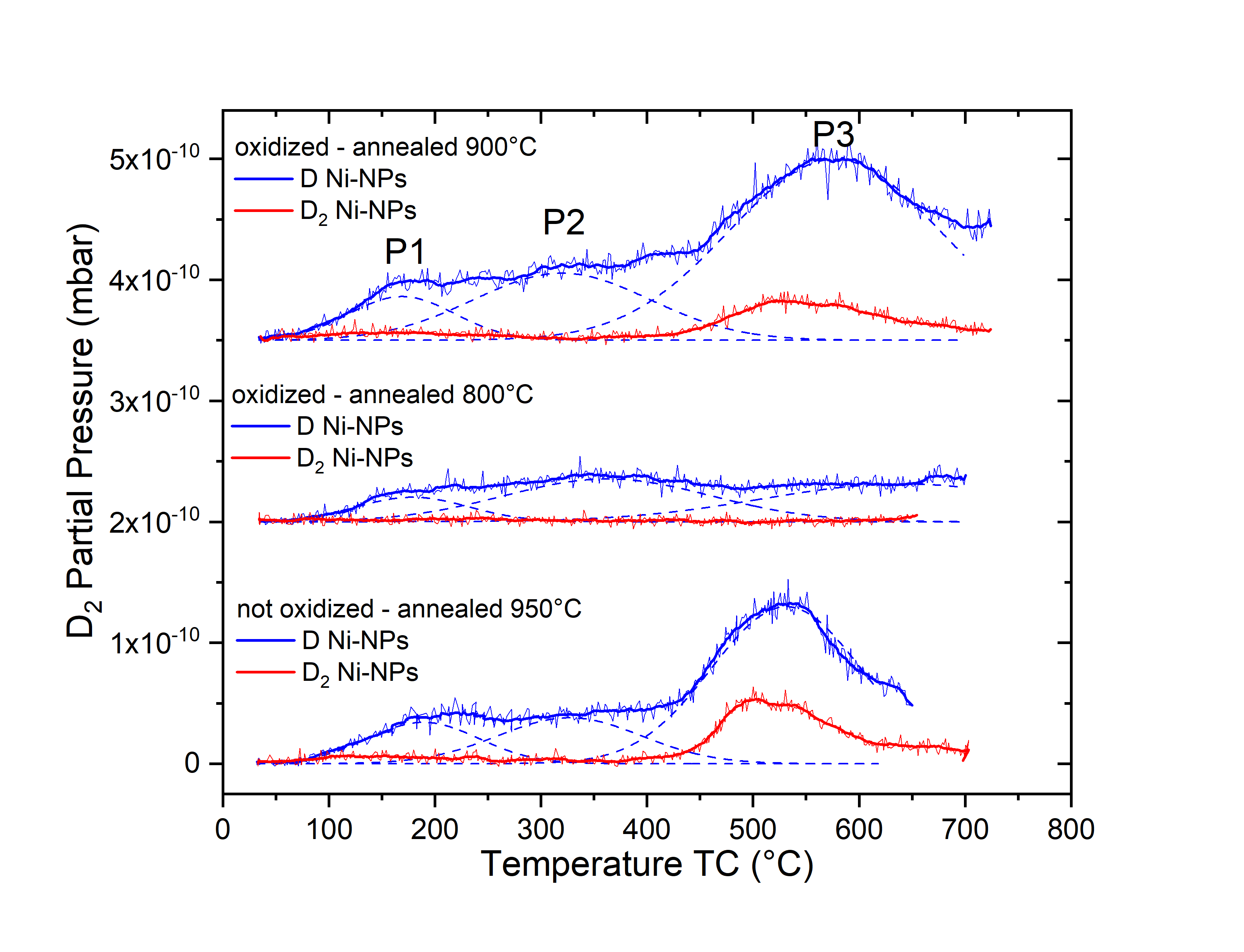}
    \caption{Comparison of TDS spectra (for D\(_2\) and D  exposure, red and blue curves, respectively) of a Ni-3DG sample after annealing at 950 \textdegree C (lower spectra), after 48 h of atmospheric exposure and 800 \textdegree C annealing (central spectra), and after further 900 \textdegree C annealing (upper spectra). Peak fits are reported for D exposure only. Samples were exposed to hydrogen for 5 minutes at \(1 \times 10^{-7}\) mbar prior to TDS, with constant heating rate of 4\,K\,s\(^{-1}\). The curves after air exposure were shifted by 2.0$\times 10^{-10}$ and 3.5$\times 10^{-10}$ mbar for better visualization.}
    \label{fig:5}
\end{figure}

In the Ni-functionalized sample, a marked increase in the P2 desorption peak centered around 350 \textdegree C is observed. Previous research indicates that desorption in this temperature region is related to defect density in the graphene lattice \cite{Macili2023,Delfino2024}. In contrast, the intensities of the lower (P1) and higher (P3) temperature desorption peaks remain relatively similar between the two samples.

The impact of oxidative processes on the hydrogen storage efficiency has been evaluated by exposing an annealed Ni-3DG sample to atmosphere. Both atomic (D) and molecular (D$_2$) deuterium exposures are performed for each hydrogenation experiment. The results of sequential oxidation/annealing/hydrogen-exposure and desorption experiments are displayed in Figure~\ref{fig:5}. The two lower TDS spectra (red and blue curves) show the hydrogen desorption peaks of Ni-functionalized 3D-Graphene after annealing at 950 \textdegree C, red for D$_2$ and blue for D. The blue curve shows a desorption spectrum close to that already shown in Figure~\ref{fig:2}. The exposure to molecular hydrogen produces a TDS spectrum (red curve) where only the (P3) peak centered around 550 \textdegree C is visible. The sample was then removed from the UHV chamber, exposed to air for 48 h, and reintroduced in the UHV chamber. The sample was annealed at 800 \textdegree C and newly exposed to atomic and molecular hydrogen (central blue and red curves in Figure~\ref{fig:5}). The oxidation effect is well evident. Compared with spectra measured before oxidation, all  peaks are suppressed after exposure to D$_2$. In particular, the largest peak (P3) is completely suppressed in the D$_2$ exposure experiment and  reduced in the D exposure experiment. Interestingly, the (P2) peak present after exposure to D is preserved in intensity with respect to the spectrum before  exposure to air.

The sample was then further annealed at 900 \textdegree C and exposed again to hydrogen (upper blue and red curves). Upon atomic hydrogen exposure, the high-temperature desorption peak (P3) as well the (P1) peak recover, indicating a recurring yet reversible effect of oxidation. Moreover, the (P2) peak intensity increases by about 85\% compared to before  exposure to air. The area under the peaks after D exposure is reported in Table~\ref{table:O2_peaks_comparison}, allowing a quantitative evaluation of the oxidation impact. A similar recovery is visible in the TDS spectrum after molecular hydrogen exposure.

Previously, the effect of graphene curvature on the atomic hydrogen adsorption has been investigated showing that the preferred binding sites correspond to convex surfaces, closer to a sp$^3$ configuration \cite{Goler2013}. Here, the presence of intercalated Ni islands would provide a larger amount of suitable sites for hydrogen binding at the edges of the islands. This feature is not affected by oxidation. Therefore, the observation that the mid-temperature peak (P2) remains largely unaffected by oxidation is consistent with the presence of intercalated Ni islands which provide adsorption sites protected from oxygen. The intercalated areas increase with the number of annealing cycles \cite{Vlamidis2025}, therefore, the notable increase in the final spectrum (+85\%) is a further support of the Ni intercalation. Moreover, it suggests that the hydrogen storage capacity is maintained and further enhanced regardless of the oxidative cycle. We have not performed a structural characterization of the defect density and surface curvature, however, the behavior of the P2 desorption peak is accordingly consistent with our hypothesis of intercalation. Indeed, the intercalated Ni islands provide adsorption sites for hydrogenation, likely due to the induced curvature of the surface at the edges of intercalated islands, not affected by the exposure to oxygen. Moreover, the intercalated areas increase upon successive annealing, leading to an increase in the P2 peak intensity.

\begin{table}[h]
    \centering
    \caption{Comparison of peak areas, after D exposure, after 950 \textdegree C annealing, O\textsubscript{2} exposure and 800 \textdegree C annealing, and after   900 \textdegree C annealing. Variation in the values is given with respect to the 950 \textdegree C reference values.}
    \label{table:O2_peaks_comparison}
    \resizebox{\columnwidth}{!}{%
        \begin{tabular}{|c|c|c|c|}
            \hline
            \textbf{Spectrum} & \textbf{P1 (mbar × K)} & \textbf{P2 (mbar × K)} & \textbf{P3 (mbar × K)} \\ 
            \hline
            950 \textdegree C Reference & $4.52 \times 10^{-9}$ & $6.47 \times 10^{-9}$ & $2.30 \times 10^{-8}$ \\ 
            O\textsubscript{2} 800 \textdegree C & $2.40 \times 10^{-9}\ (-46\%)$ & $8.12 \times 10^{-9}\ (+25\%)$ & $9.66 \times 10^{-9}\ (-58\%)$ \\ 
            O\textsubscript{2} 900 \textdegree C & $6.5 \times 10^{-9}\ (+43\%)$ & $1.2 \times 10^{-8}\ (+85\%)$ & $2.8 \times 10^{-8}\ (+21\%)$ \\ 
            \hline
        \end{tabular}%
    }
\end{table}

\section{Conclusions}

Our results collectively highlight how temperature-driven processes and nickel functionalization enhance hydrogen storage in 3D-Graphene. The increased hydrogen uptake is driven by the substrate annealing temperature (independently from the Ni functionalization) and by the Ni functionalization. Higher annealing temperatures correlate with the removal of surface silicon oxides, creating additional adsorption sites. Indeed, TDS measurements for pristine 3D-Graphene annealed at 900 \textdegree C exhibit a stronger desorption peak (P3) around 550 \textdegree C compared to those annealed at 800 \textdegree C, reflecting an improvement in hydrogen storage capacity. The XPS analysis shows a decrease in SiO$_2$ signal intensity for higher annealing temperatures, confirming that oxide species are progressively desorbed and suggesting that desorption makes  binding sites for hydrogen available. Indeed, the SiO$_2$ thermal decomposition temperature matches very well with the observed increase in hydrogen uptake between 800 \textdegree C and 900 \textdegree C. 

In Ni-functionalized samples, TDS spectra reveal a notably intensified mid-tem\-per\-a\-ture peak, consistent with curvature-mediated adsorption reported in literature \cite{Goler2013}. This increase is accompanied by morphological and chemical changes of the Ni nanoparticles, as observed by SEM and XPS. Once the capping layer of the Ni NPs decomposes, nickel eventually intercalates beneath the graphene, leading to an attenuated Ni 2p XPS signal after high annealing temperatures. This intercalation correlates with the stability and persistence of the Ni and thus the mid-temperature desorption feature.

Oxidation experiments with Ni-functionalized samples further demonstrate that the high-temperature desorption peak (P3) diminishes upon exposure to atmospheric oxygen but reappears after renewed annealing at 900 \textdegree C, indicating a reversible process. By contrast, the P2 peak remains approximately unaffected by oxidation, and is even enhanced after multiple cycles, underscoring its robustness. All in all, these observations suggest that Ni-functionalized 3D-Graphene retains enhanced hydrogen storage performance even after repeated oxidation and regeneration steps, thereby offering a promising platform for advanced solid-state hydrogen storage applications.

\bibliography{Bibliography}

\end{document}